\documentclass[%
 reprint,
 amsmath,amssymb,
 aps,
]{revtex4-1}

\usepackage{graphicx}
\usepackage{dcolumn}
\usepackage{bm}

\newcommand{\bea}{\begin{eqnarray}\displaystyle}
\newcommand{\eea}{\end{eqnarray}}
\newcommand{\nn}{\nonumber}

\newcommand{\figref}[1]{Fig.~\protect\ref{#1}}

{\setlength{\fboxsep}{15pt}
\setlength{\mylength}{\linewidth}%
\addtolength{\mylength}{-2\fboxsep}%
\addtolength{\mylength}{-2\fboxrule}%
\Sbox
\minipage{\mylength}%
\setlength{\abovedisplayskip}{0pt}%
\setlength{\belowdisplayskip}{0pt}%
\equation}%
{\endequation\endminipage\endSbox
\[\fbox{\TheSbox}\]}
\begin{document}


\title{Brane Webs and Random Processes}
\author{Amer Iqbal$^{a,b,c,d}$, Babar A. Qureshi$^{a}$, Khurram Shabbir$^{e}$, Muhammad A. Shehper$^{a}$}
 \vskip 2cm
 \affiliation{$^{a}$Department of Physics, School of Science \& Engineering, LUMS, Lahore, Pakistan}
\affiliation{$^{b}$Department of Mathematics, School of Science \& Engineering, LUMS, Lahore, Pakistan}

 \affiliation{$^{c}$Abdus Salam School of Mathematical Sciences, GC University, Lahore, Pakistan}
\affiliation{$^{d}$Center of Mathematical Sciences and Applications, Harvard University
1 Oxford Street, Cambridge, MA 02138, USA}
\affiliation{$^{e}$Department of Mathematics, Government College University, Lahore, Pakistan.}%


\date{\today}

\begin{abstract}
We study $(p,q)$ 5-brane webs dual to certain $N$ M5-brane configurations and show that the partition function of these brane webs gives rise to cylindric Schur process with period $N$. This generalizes the previously studied case of period $1$. We also show that open string amplitudes corresponding to these brane webs are captured by the generating function of cylindric plane partitions with profile determined by the boundary conditions imposed on the open string amplitudes.
\end{abstract}

\maketitle

\tableofcontents

\section{Introduction}
Topological strings on toric Calabi-Yau threefolds provide an interesting set of examples which are quite well understood. The topological vertex \cite{Aganagic:2003db} and refined topological vertex \cite{Iqbal:2007ii} formalism provide an exact solution of the topological string partition functions in the unrefined and the refined case respectively.  The topological vertex formalism reduces the calculation of the topological string partition function to sums over functions of Young diagrams. In case the toric Calabi-Yau threefold gives rise to gauge theory, via geometric engineering \cite{Katz:1996fh}, the Young diagrams appearing in the partition functions of topological string can be directly be related to Young diagrams in the Nekrasov's instanton calculus \cite{Nekrasov:2002qd} which label the fixed points on the instanton moduli spaces. It has been shown that in certain cases these gauge theory partition functions can be thought of as sums over probability measures on the set of Young diagrams \cite{Nekrasov:2003rj}. In \cite{Iqbal:2008ra} it was shown that the partition function of  5D ${\cal N}=1$ $U(1)$ gauge theory with an adjoint hypermultiplet compactified on a circle gives a probability measure of a random process studied by Borodin in \cite{borodin} called the periodic Schur process with period one. This gauge theory also arises from mass deformation of an M5-brane with a transverse direction compactified to a circle \cite{mstrings}. In this short note we show that the probability measure associated with periodic Schur process with period $N$ is given by the gauge theory partition function arises from N-M5-branes threading a circle. The gauge theory this configuration gives rise to is the $U(1)^N$ gauge theory with bifundamental matter. 

The paper is organized as follows. In section 2 we discuss the various generalizations of the Plancherel measure which arise from the $N=1$ case. In section 3 we discuss the M5-brane configuration and the corresponding dual $(p,q)$ brane web configuration which corresponds to the period $N$ periodic Schur process. In this section we also calculate the topological string partition function, using the topological vertex formalism, of the Calabi-Yau threefold dual to the brane web and show that it is gives the periodic Schur process of period $N$. In section 4 we show that the periodic Schur process with non-trivial profile corresponds to certain open topological string amplitudes. In section 5 we present our conclusions and future directions.

\section{Probability measures and gauge theories}
In \cite{Nekrasov:2003rj} it was shown that the partition function of the four dimensional $U(1)$ gauge theory with adjoint hypermultiplet can be written as sum over Young diagram of a function which can be thought as a probability measure on the set of Young diagrams. In this section we study some generalizations of the Plancherel measure which follow from 5D $U(1)$ and 6D $U(1)$ gauge theory with adjoint hypermultiplet.

\subsection{Plancherel Measure}
\begin{figure}[h]
  \includegraphics[width=0.75in]{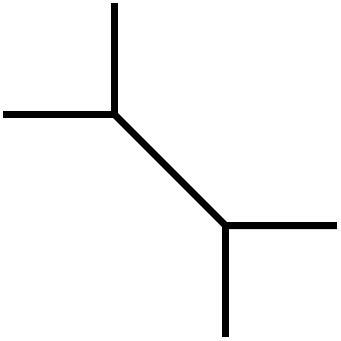}
  \caption{$(p,q)$ brane web dual to the resolved conifold.}\label{braneweb1}
\end{figure}

The case of the Plancherel measure was shown to follow from the four dimensional ${\cal N}=2$ $U(1)$ gauge theory \cite{Nekrasov:2003rj}. This four dimensional gauge theory can be obtained from the five dimensional theory by circle compactfication, the 5D theory can be geometrically engineered by M-theory compactification on the resolved conifold. The 5D theory can also be realized on a $(p,q)$ 5-brane web, shown in \figref{braneweb1}, in type IIB string theory \cite{Aharony:1997bh}.

Using the refined topological vertex formalism to the brane web in \figref{braneweb1} the gauge theory partition function is given by,
\bea
Z&=&\sum_{\lambda}Q^{|\lambda|}\,s_{\lambda}(t^{-\rho})s_{\lambda^t}(q^{-\rho})\\\nn
&=&\sum_{\lambda}Q^{|\lambda|}\,\frac{t^{\frac{||\lambda^t||^2}{2}}q^{\frac{||\lambda||^2}{2}}}{\prod_{(i,j)\in \lambda}(1-q^{h_{\lambda}(i,j)})(1-t^{h_{\lambda}(i,j)})}
\eea
From this we see that
\bea
P_{\lambda}(t,q,Q)=\frac{1}{Z}\,\frac{Q^{|\lambda|}\,t^{\frac{||\lambda^t||^2}{2}}q^{\frac{||\lambda||^2}{2}}}{\prod_{(i,j)\in \lambda}(1-q^{h_{\lambda}(i,j)})(1-t^{h_{\lambda}(i,j)})}
\eea
defines a probability measure on the set of partitions for $q,t, Q<1$ which is a generalization of the poissonized Plancherel measure. To see this take $q=t=e^{\epsilon}$, $Q=\epsilon^{2}\,\Lambda$ and consider the limit $\epsilon\mapsto 0$,
\bea\nonumber
\lim_{\epsilon\mapsto 0}P_{\lambda}(e^{\epsilon},e^{\epsilon},\epsilon^{2}\Lambda)&=&e^{-\Lambda}\,\Lambda^{|\lambda|}\prod_{(i,j)\in \lambda}h_{\lambda}(i,j)^{-2}\\
&=&e^{-\Lambda}\Lambda^{|\lambda|}\,\,\Big(\frac{\text{dim}\lambda}{|\lambda|!}\Big)^2.
\eea
The above is precisely the poissonized Plancherel measure \cite{borodin}. In the above we have used
\bea\nonumber
\lim_{\epsilon\mapsto 0}Z(e^{\epsilon},e^{\epsilon},\epsilon^{2}\Lambda)&=&\lim_{\epsilon\mapsto 0}\prod_{i,j}\Big(1-\epsilon^{2}\,\Lambda\,e^{(i+j-1)\epsilon}\Big)=e^{\Lambda}
\eea
and the hook length formula,
\bea
\text{dim}\lambda=\frac{|\lambda|!}{\prod_{(i,j)\in \lambda}h_{\lambda}(i,j)},
\eea
where $\text{dim}\lambda$ is the number of standard Young tableaus of shape $\lambda$.

In this case if we consider $t=e^{\beta\,\epsilon}$ and then take $\epsilon\mapsto 0$ we do not get any $\beta$ deformed measure instead just get the same measure with $\Lambda\mapsto \Lambda/\beta$.

\subsection{Nekrasov-Okounkov Measure and its $(q,t)$ Deformation}

\begin{figure}[h]
  \includegraphics[width=2in]{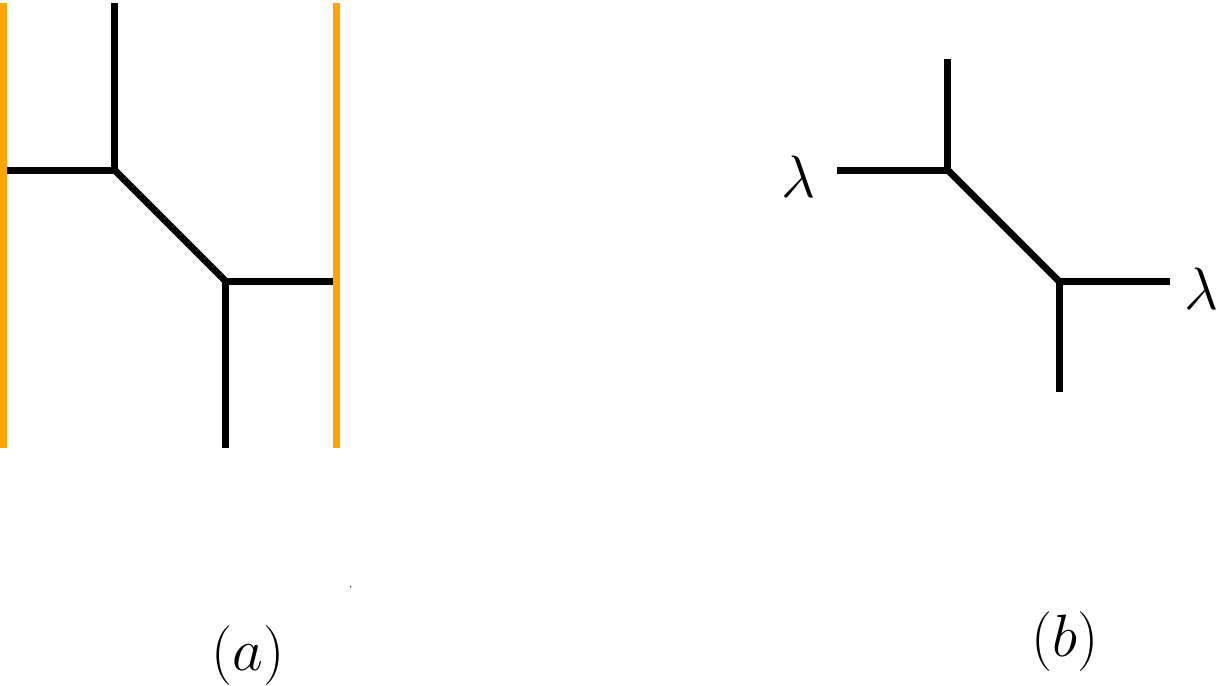}
  \caption{(a) $(p,q)$ brane web of the mass deformed theory. The brane web is now on $\mathbb{R}\times S^1$. (b) The horizontal lines are glued together and has the same partition associated with it.}\label{braneweb2}
\end{figure}

The 5D $U(1)$ gauge theory can be obtained from 5D $U(1)$ gauge theory with an adjoint hypermultiplet in the limit that the mass of the adjoint goes to infinity. The mass deformed 5D theory is realized by the brane web shown in \figref{braneweb2}.

The partition function is given by
\bea
Z_{5D}&=&Z_{4D}\,\sum_{\lambda}\,{\cal M}_{\lambda}\\\nonumber
{\cal M}_{\lambda}&=&Q_{\rho}^{|\lambda|}\,\prod_{s\in \lambda}\frac{(1-\widehat{Q}\,q^{\ell(s)+1}\,t^{a(s)})(1-\widehat{Q}^{-1}q^{\ell(s)}\,t^{a(s)+1})}{(1-q^{\ell(s)}t^{a(s)+1})(1-q^{\ell(s)+1}t^{a(s)})}
\eea
where $\widehat{Q}=Q\sqrt{\frac{t}{q}}$ and  $a(i,j)=\nu^{t}_{j}-i\,,\,\,\,\,\ell(i,j)=\nu_i-j$. If we define $W=\sum_{\lambda}{\cal M}_{\lambda}$ then
\bea
{\cal P}_{\lambda}(t,q,Q,Q_{\rho})=\frac{1}{W}\,{\cal M}_{\lambda}
\eea
defines a probability measure on the set of partitions which is a $(q,t)$ deformation of the Nekrasov-Okounkov measure \cite{Nekrasov:2003rj}.
\bea\nonumber
\lim_{\epsilon\mapsto 0}{\cal P}_{\lambda}(e^{\epsilon},e^{\epsilon},e^{\mu\,\epsilon},Q_{\rho})=\frac{Q_{\rho}^{|\lambda|}\,\prod_{s\in \lambda}\frac{h(s)^{2}-\mu^2}{h(s)^2}}{\prod_{k=1}^{\infty}(1-Q_{\rho}^{k})^{\mu^{2}-1}}
\eea

\bea\nonumber
\lim_{t\mapsto 0}{\cal P}_{\lambda}(e^{t\epsilon_{2}},e^{t\epsilon_{1}},e^{t\,\mu},Q_{\rho})=\frac{Q_{\rho}^{|\lambda|}\,\prod_{s\in \lambda}f(s,\epsilon_1,\epsilon_2,\mu)}{\lim_{t\mapsto 0}W}
\eea
\bea\nonumber
f(s,\epsilon_{1,2},\mu)=\frac{(a(s)+1+\beta\,\ell(s)-\frac{\mu}{\epsilon_{1}})(a(s)+\beta(\ell(s)+1)+\frac{\mu}{\epsilon_{1}})}{(a(s)+1+\beta\ell(s))(a(s)+\beta(\ell(s)+1))}
\eea
where $\beta=\frac{\epsilon_{2}}{\epsilon_{1}}$ and
\bea
\lim_{t\mapsto 0}W=\prod_{k=1}^{\infty}(1-Q_{\rho}^k)^{\frac{(\mu-\epsilon_{1})(\mu+\epsilon_{2})}{\epsilon_{1}\epsilon_{2}}}
\eea
which gives the $\beta$ deformed Nekrasov-Okounkov measure.

\subsection{Generalization of Nekrasov-Okounkov Measure Involving Theta Function}
The further compactification of the web shown in \figref{braneweb2} by changing the space on which the web lives from a cylinder to a torus gives a generalization of the Nekrasov-Okounkov measure involving theta functions.

\begin{figure}[h]
  \includegraphics[width=1.65in]{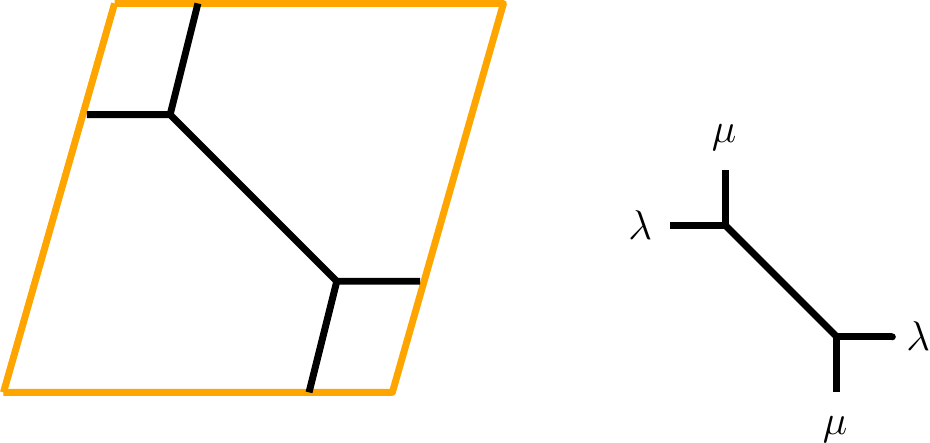}\\
\caption{Brane web on $T^2$.}\label{compactweb}
\end{figure}

The partition function in this case is given by
\bea\nonumber
Z_{6D}&=&Z_{5D}\sum_{\lambda}\mathbb{M}_{\lambda}\\\nonumber
\mathbb{M}_{\lambda}&=&Q_{\rho}^{|\lambda|}\,\prod_{s\in \lambda}\frac{\theta_{1}(m+z(s))\theta_{1}(m+z(s)-\epsilon_{+})}{\theta_{1}(z(s))\theta_{1}(z(s)-\epsilon_{+})}
\eea
where $z(s)=(\ell(s)+1)\epsilon_{1}-a(s)\epsilon_{2}$ and $\epsilon_{+}=\epsilon_{1}+\epsilon_{2}$.
\bea
\mathbb{P}_{\lambda}=\mathbb{W}^{-1}\,\mathbb{M}_{\lambda}
\eea
where
\bea\nonumber
\mathbb{W}=\frac{Z_{6D}}{Z_{5D}}\,.
\eea
$\bullet$ $m\mapsto 0$ gives uniform measure $Q_{\rho}^{|\lambda|}$.\\
$\bullet$ $\tau\mapsto i\infty$ gives the $(q,t)$ deformed Nekrasov-Okounkov measure.

\section{Periodic Schur Process and Brane Webs}

\subsection{Periodic Schur process of period $N$}
In this section we review, following \cite{borodin}, the periodic Schur process and the probability measure associated with it.

We denote by $\mathbb{Y}$ the set of Young diagrams. Then \cite{borodin} defines the periodic Schur process of period $N$ to be a random process defined on $\mathbb{Y}^{2N}$ which
assigns to the set of partitions $\{\nu^{0},\mu^{1},\nu^{1},\mu^{2},\cdots, \nu^{N-1},\mu^{N}\}\in \mathbb{Y}^{2N}$ the weight,
\bea
Q_{\bullet}^{|\nu^{0}|}\Big(\prod_{a=0}^{N-1}s_{\nu^{a}/\mu^{a+1}}(x_{a+1})s_{\nu^{a+1}/\mu^{a+1}}(y_{a+1})\Big)/Z_{N}
\eea
where $\nu^{N}=\nu^{0}$ and $Z_{N}$ is the normalization which is also the partition function
of this random process,
\bea\label{schurprocess}
Z_{N}&=&\sum_{\nu^{a},\mu^{a}}Q_{\bullet}^{|\nu^{0}|}\prod_{a=0}^{N-1}s_{\nu^{a}/\mu^{a+1}}(x_{a+1})\\\nonumber
&&s_{\nu^{a+1}/\mu^{a+1}}(y_{a+1})\,.
\eea
We will see in the next section that for a particular specialization $x_{a}$ and $y_{a}$ the above partition function of the periodic Schur process will be exactly the partition function of a configuration of $N$ M5-branes.

\subsection{$(p,q)$ 5-brane webs and M5-branes}

The case of periodic Schur process of period $1$ was discussed in detail in \cite{Iqbal:2008ra}. The brane web for this case is shown in \figref{braneweb2}. An M5-brane realization of this theory was studied in \cite{mstrings} where it was shown that the $(p,q)$ 5-brane web of \figref{braneweb2}is dual to a configuration in which we have a single M5-brane compactified on a circle. The space transverse to the M5-brane is $\mathbb{R}^5$ and there is $U(1)$ action on the $\mathbb{C}^2\subset \mathbb{R}^5$ one goes around the transverse circle,
\bea\label{mass}
U(1): (w_{1},w_{2})\mapsto (e^{2\pi i m}w_{1},e^{-2\pi im}w_{2})\,.
\eea

This M5-brane configuration is, however, dual to another configuration in which the M5-brane is not wrapped but there is a circle transverse to the M5-brane. When the M5-brane is wrapped on a circle the massive modes come from Kaluza-Klein reduction of the 6D free tensor multiplet on a circle as shown in \cite{mstrings}. In the case when the M5-brane is not wrapped but there is a circle transverse to it the massive modes come from the M2-brane starting and ending of the M5-brane and wrapping the transverse circle.

\begin{figure}[h]
  \includegraphics[width=3.4in]{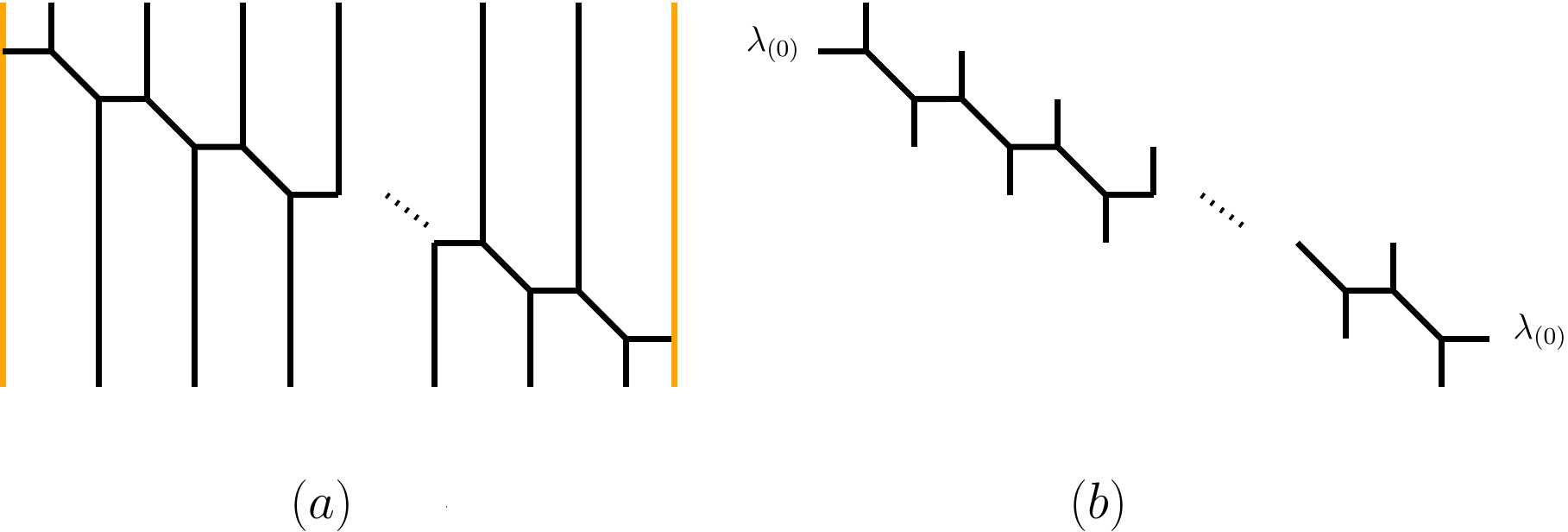}
  \caption{(a) $(p,q)$ brane web of the mass deformed theory. The brane web is now on $\mathbb{R}\times S^1$. (b) The horizontal lines are glued together and has the same partition associated with it.}\label{braneweb4}
\end{figure}

Now consider the brane configuration in which we have multiple coincident M5-branes and a circle transverse to them. We can separate the M5-branes on that circle and mass deform the configuration as given in Eq.(\ref{mass}). This brane configuration is dual to a $(p,q)$ 5-brane web given in \figref{braneweb4}(a). This brane web gives rise to $U(1)^N$ gauge theory with bifundamental matter.

\begin{figure}[h]
  \includegraphics[width=3in]{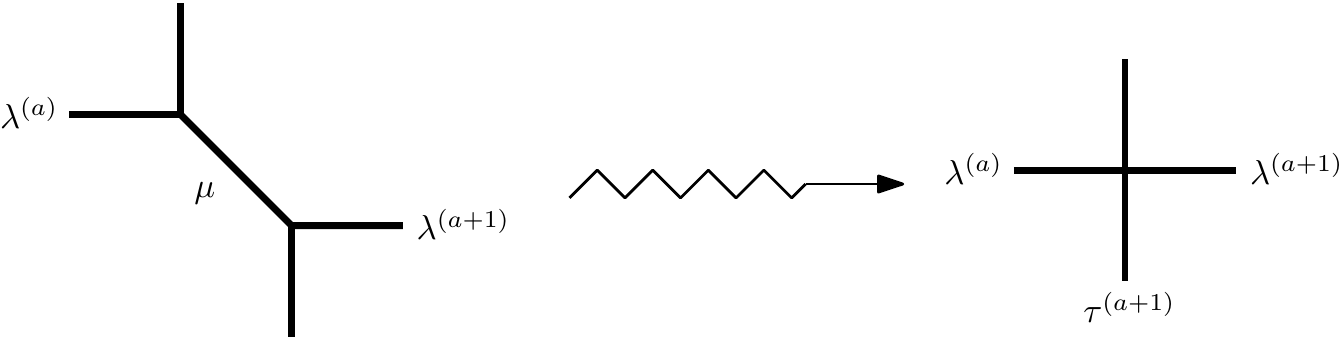}\\
\caption{The open string amplitude which is building block of the topological string partition function. After summing over $\mu$ an auxiliary partition $\tau$ appears which is same as the one appearing in the definition of the periodic Schur process.}\label{schur2}
\end{figure}

 The instanton partition function of this gauge theory can be calculated using the refined topological vertex formalism. We denote the topological string partition function of this Calabi-Yau threefold by $Z_{N}$ and it is given by
\bea\label{PF1}
Z_{N}:=\sum_{\vec{\lambda}}\prod_{a=0}^{N-1}\Big[(-Q_{a})^{|\lambda_{(a)}|}\,W_{\lambda_{(a)}\lambda_{(a+1)}}\Big]\,,
\eea
where
\bea\nonumber
W_{\lambda_{(a)}\lambda_{(a+1)}}=\sum_{\mu}(-Q_{m})^{|\mu|}\,C_{\lambda^{t}_{(a)}\,\mu\emptyset}(t, q)\,C_{\lambda_{(a+1)}\mu^{t}\emptyset}(q, t)
\eea
and 
\bea
C_{\lambda\mu\emptyset}(t,q)&=&\Big(\frac{q}{t}\Big)^{\frac{|\lambda|+||\mu||^2-|\mu|}{2}}\,t^{\frac{\kappa(\mu)}{2}}\widetilde{Z}_{\nu}(t,q)\\\nonumber
&&\times\sum_{\eta}\Big(\frac{q}{t}\Big)^{\frac{|\eta|}{2}}\,s_{\lambda^{t}/\eta}(t^{-\rho})\,s_{\mu/\eta}(q^{-\rho})
\eea
is the refined topological vertex. For notation see Appendix A. The length of the slanted lines in \figref{braneweb4} are all equal to $m$ and we have defined $Q_{m}=e^{-m}$, similarly the length of the horizontal lines is $T_{a}$ and we have defined $Q_{a}=e^{-T_{a}}$ such that $-log(Q_{a}Q_{m})$ is the distance between the two vertical lines. In Eq.(\ref{PF1}) $W_{\lambda\nu}$ is the open string amplitude corresponding the brane configuration shown in \figref{schur2} below and is given by

\bea\nonumber
W_{\lambda\nu}&=&\Big(\frac{q}{t}\Big)^{\frac{|\lambda|-|\nu|}{2}}\sum_{\eta_{1},\eta_{2}}\Big(\frac{q}{t}\Big)^{\frac{|\eta_{1}|-|\eta_{2}|}{2}}s_{\lambda/\eta_{1}}(t^{-\rho})\,s_{\nu^{t}/\eta_{2}}(q^{-\rho})\\\label{p1}&&\times \sum_{\mu}
(-Q_{m})^{|\mu|}s_{\mu/\eta_{1}}(q^{-\rho})s_{\mu^{t}/\eta_{2}}(t^{-\rho})
\eea

Using the identity
\bea\label{id1}
\sum_{\lambda}s_{\lambda^{t}/\eta}(x)s_{\lambda/\sigma}(y)&=&\prod_{i,j}(1+x_{i}y_{j})\,\\\nonumber
&&\times \sum_{\tau}s_{\sigma^{t}/\tau}(x)\,s_{\eta^{t}/\tau^{t}}(y)
\eea
the sum over $\mu$ in Eq.(\ref{p1}) can be carried out and we get,
\bea\label{p2}
&&W_{\lambda\nu}=\Pi(Q_{m})\,\Big(\frac{q}{t}\Big)^{\frac{|\lambda|-|\nu|}{2}}\sum_{\eta_{1},\eta_{2}}\Big(\frac{q}{t}\Big)^{\frac{|\eta_{1}|-|\eta_{2}|}{2}} s_{\lambda/\eta_{1}}(t^{-\rho})\\\nonumber&&\,s_{\nu^{t}/\eta_{2}}(q^{-\rho})\sum_{\tau}(-Q_{m})^{|\tau|}s_{\eta^{t}_{1}/\tau}(-Q_{m}t^{-\rho})s_{\eta^{t}_{2}/\tau^t}(-Q_{m}q^{-\rho})
\eea
where
\bea
\Pi(x)=\prod_{i,j=1}^{\infty}\Big(1-x\,q^{-\rho_{i}}\,t^{-\rho_{j}}\Big)\,.
\eea

Using the following properties of the skew-schur functions,
\bea\label{id2}
s_{\lambda^t/\sigma^t}(q^{-\rho})=s_{\lambda/\sigma}(-q^{\rho})\\\nonumber
\sum_{\eta}s_{\lambda/\eta}(x)\,s_{\eta/\sigma}(y)=s_{\lambda/\sigma}(x,y),
\eea
we get,
\bea
W_{\lambda\nu}&=&\Big(\frac{q}{t}\Big)^{\frac{|\lambda|-|\nu|}{2}}\,(-1)^{|\nu|}\,\Pi(Q_{m})\,\widetilde{W}_{\lambda\nu}\\\nonumber
\widetilde{W}_{\lambda\nu}&=&\sum_{\tau}Q_{m}^{|\tau|}s_{\lambda/\tau}({\bf a})\,s_{\nu/\tau}({\bf b})\,.
\eea
In the above equation,
\bea
{\bf a}=\{Q_{m}\sqrt{\frac{q}{t}}t^{\rho},t^{-\rho}\}, {\bf b}=\{Q_{m}\sqrt{\frac{t}{q}}q^{-\rho},q^{\rho}\}
\eea
The partition function in Eq.(\ref{PF1}) can now be written as,
\bea\label{PF}
Z_{N}:=\Pi(Q_{m})^{N}\,\sum_{\vec{\lambda}}\prod_{a=0}^{N-1}\Big[Q_{a}^{|\lambda_{(a)}|}\,\widetilde{W}_{\lambda_{(a)}\lambda_{(a+1)}}\Big]\,.
\eea
The partition function $\widehat{Z}_{N} := Z_{N}/\Pi(Q_{m})^N$ is thus given by,
\bea\label{PF2}
\widehat{Z}_{N}&:=&\sum_{\vec{\lambda}}\prod_{a=0}^{N-1}\Big[Q_{a}^{|\lambda_{(a)}|}\,\widetilde{W}_{\lambda_{(a)}\lambda_{(a+1)}}\Big]\\\nonumber
&=&\sum_{\vec{\lambda}, \vec{\tau}} \prod_{a=0}^{N-1} Q_a^{|\lambda_{(a)}|} Q_m^{|\tau_{(a+1)}|} s_{\lambda_{(a)}/\tau_{(a+1)}}({\bf a})\,s_{\lambda_{(a+1)}/\tau_{(a+1)}}({\bf b})  \\\nonumber
&=&\sum_{\vec{\lambda}, \vec{\tau}} Q_{\rho}^{|\lambda_{(0)}|}\,\prod_{a=0}^{N-1}s_{\lambda_{(a)}/\tau_{(a+1)}}({\bf x}_{a+1})\,s_{\lambda_{(a+1)}/\tau_{(a+1)}}({\bf y}_{a+1})\,,
\eea
where we have defined new variables: 
\bea \nonumber 
Q_{\rho} = \prod\limits_{a=0}^{N-1} (Q_a Q_m)\quad ,\quad Q_{1,a+1}=(Q_{1}Q_{2}\cdots Q_{a})Q_{m}^{a}\eea
 and
\bea
{\bf x}_{a+1}&=& Q_m^{-1/2} Q_{1,a+1}\,{\bf a}\,, \\ \nonumber
{\bf y}_{a+1}&=& Q_m^{-1/2} Q_{1,a+1}^{-1}\,{\bf b}\,. \nonumber
\eea
$-log(Q_{\rho})$ is precisely the circumference of the circle. In writing last line of Eq.(\ref{PF2}) we have used the following identity,
\bea\nonumber
Q_{m}^{\sum_{a=0}^{N-1}|\lambda_{(a)}|}
\prod_{a=0}^{N-1}Q_{a}^{|\lambda_{(a)}|}&=&Q_{\rho}^{|\lambda_{(0)}|}\prod_{a=0}^{N-1}Q_{1,a+1}^{|\lambda_{(a)}|-|\tau_{(a+1)}|}\,\\\nonumber
&&\times \prod_{a=0}^{N-1}(Q_{1,a+1}^{-1})^{|\lambda_{(a+1)}|-|\tau_{(a+1)}|}\,.
\eea
We can describe the partition function graphically by associating the partition $\tau_{a}$ with each M5-brane and partitions $\lambda_{a}$ with the interval between the M5-branes as shown in \figref{schur1} below.

\begin{figure}[h]
  \includegraphics[width=3in]{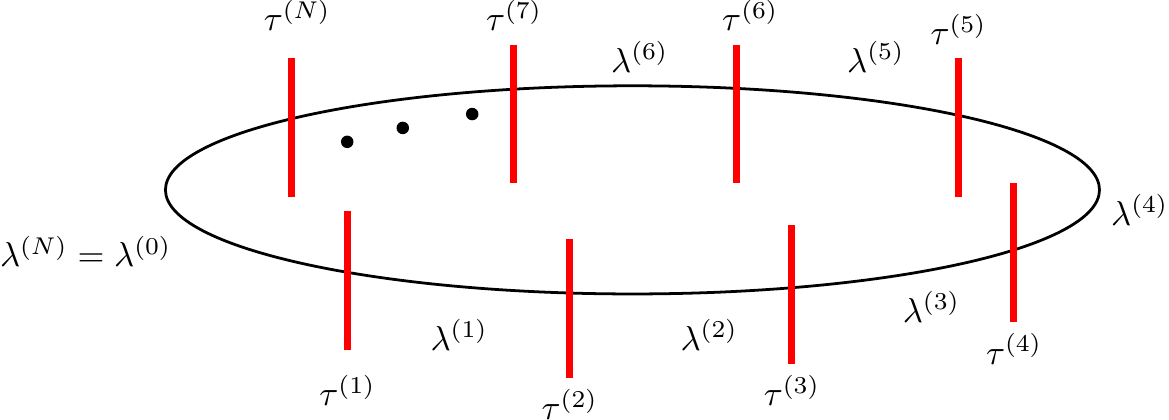}\\
\caption{After summing over the partitions associated with the slanted lines auxiliary partitions $\tau_{a}$ appear sandwiched between $\lambda_{a-1}$ and $\lambda_{a}$. The periodic Schur process is defined in terms of the set $\{\lambda_{a},\tau_{a}\}$.}\label{schur1}
\end{figure}

From last line of Eq.(\ref{PF2}) and Eq.(\ref{schurprocess}) we see that the partition function of this configuration of $N$ M5-branes is precisely the partition function of the periodic schur process with period $N$. In the limit $Q_{\rho}\mapsto 0$ we get $\lambda_{(0)}=\lambda_{(N)}=\emptyset$ and we get usual Schur process. Thus the usual Schur process is associated with linear configuration of M5-branes.

\section{Open String Amplitudes and Cylindric Partitions}
\begin{figure}[h]
  \includegraphics[width=3.0in]{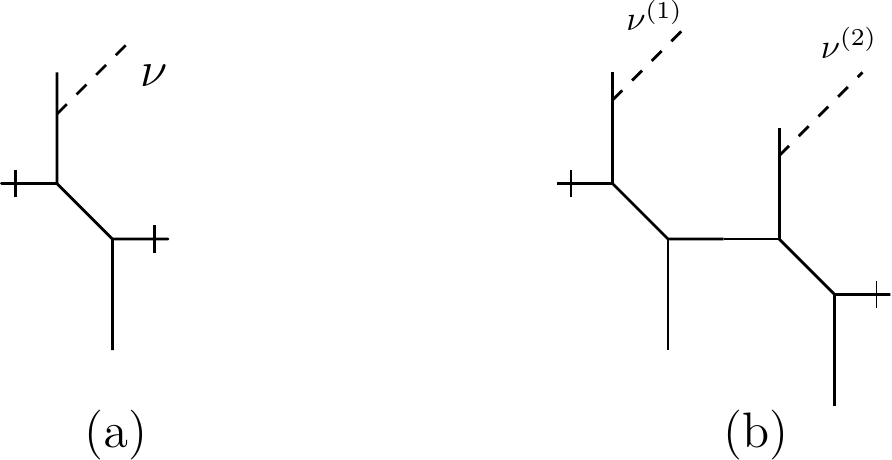}\\
\caption{Open string configurations}\label{cyl4}
\end{figure}

In this section we show that open string amplitudes corresponding to the brane configuration shown in \figref{cyl4} are also given by cylindric plane partitions \cite{GK, borodin, tingley} with non-trivial profile which captures the partitions $\nu_{(a)}$.

\begin{figure}[h]
  \includegraphics[width=3in]{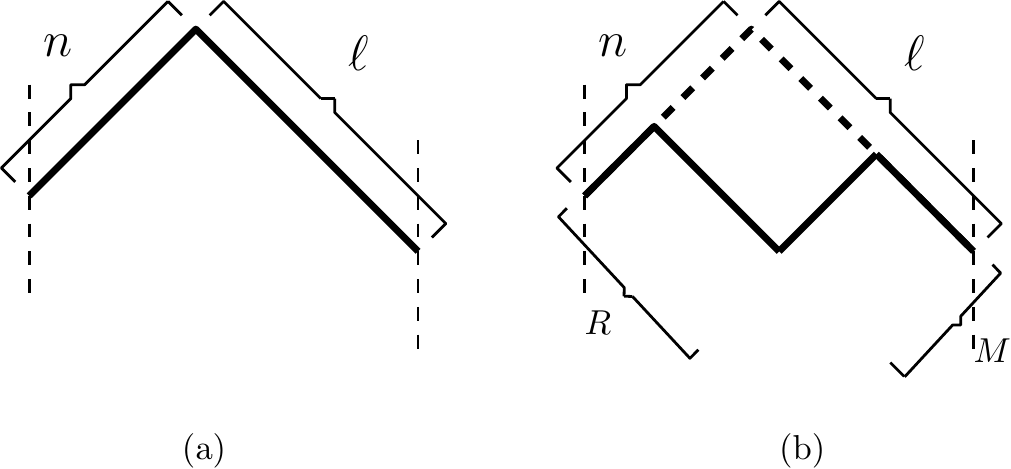}\\
  \caption{(a) Cylindric partition with empty partition as a profile. (b) Cylindric partition with profile given by a $R\times M$ partition. The vertical dotted lines indicate gluing so that the plane partitions live on a cylinder.}
  \label{cyl1}
\end{figure}

Let us begin by considering the cylindric partition with trivial profile as shown below in \figref{cyl1}(a). 

\begin{figure}[h]
  \includegraphics[width=3in]{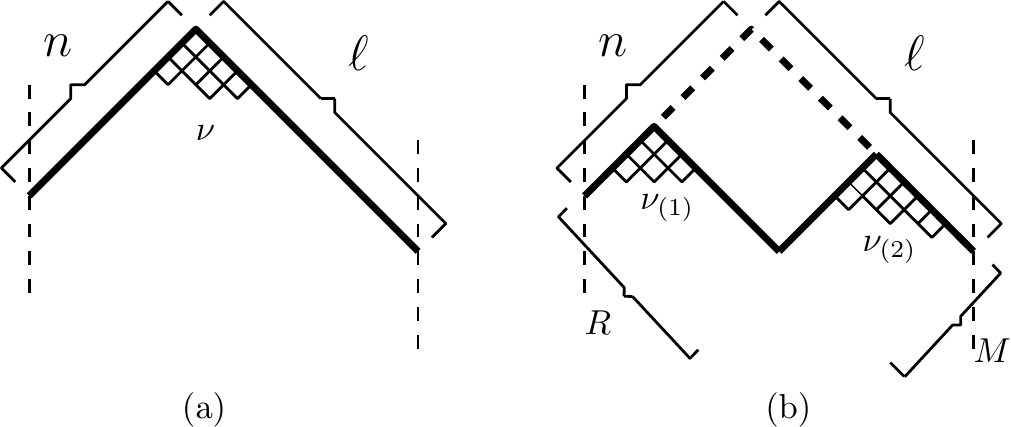}\\
\caption{(a) region in the shape of $\nu$ is excised and plane partitions are put in the remaining region. (b) Regions in the shape of $\nu_{1}$ and $\nu_{2}$ are excised and plane partitions put in the remaining region. }\label{cyl2}
\end{figure}

The generating function for cylindric plane partitions with this profile was shown in \cite{Iqbal:2008ra} to be equal to the partition function of the brane configuration in \figref{braneweb4} with $N=1$. If we change the profile to be a $M\times R$ rectangular Young diagram then the generating function of cylindric partitions with this profile is given by \cite{borodin, Iqbal:2009ki},
\bea\nonumber
\mathbb{G}^{\ell,n}_{\sigma}(q)=\prod_{s\in \sigma}(1-q^{h(s)})^{-1}\prod_{k=1}^{\infty}
(1-Q_{\rho}^{k})^{-1}\,\prod_{i,j=1}^{\ell,n}(1-Q_{\rho}^{k}q^{h(s)}),
\eea
where $Q_{\rho}=q^{n+\ell}$ and $h(i,j)=\sigma_{i}+\sigma^{t}_{j}-i-j+1$. The partition $\sigma=(\underbrace{R,R,R,R,\cdots,R}_{M-times})$.
Now we can put a non-trivial partition at each of the corners as shown in \figref{cyl2}. It was shown in \cite{Iqbal:2009ki} that the generating function associated with \figref{cyl2}(a) is precisely the open string amplitude corresponding to the brane configuration in \figref{cyl4}(a). 

If we non-trivial partition $\nu_{1}$ and $\nu_{2}$ in the two corners of \figref{cyl2}(b) then the generating function of cylindric plane partitions is given by,
\bea\nonumber
\mathbb{G}^{\ell,n}_{\sigma}(q)=\prod_{s\in \sigma}(1-q^{h(s)})^{-1}\prod_{k=1}^{\infty}
(1-Q_{\rho}^{k})^{-1}\,\prod_{i,j=1}^{\ell,n}(1-Q_{\rho}^{k}q^{h(s)})
\eea
where $Q_{\rho}=q^{n+\ell}$ and $h(i,j)=\sigma_{i}+\sigma^{t}_{j}-i-j+1$. We take the partition $\sigma=(\underbrace{\nu^{(1)}_{1}+R,\cdots,\nu^{(1)}_{\ell(\nu^{(1)})}+R,R,R,\cdots,R}_{M-times},\nu^{(2)}_{1},\cdots,\nu^{(2)}_{\ell(\nu^{(2)})})$. The above is precisely the open string amplitude associated with brane configuration shown in \figref{cyl4}(b) for,
\bea
Q_{m_{1}}=q^{R}\,,\,\,Q_{m_{2}}=q^{\ell-R}\,,\,\,Q_{1}:=q^{n-M}\,,
\eea
where $Q_{m_{1,2}}$ are the parameters associated with the slated lines and $Q_{1}$ is the parameter associated with middle horizontal line.

\section{Conclusions}
In this short note we have shown that generalizations of Nekrasov-Okounkov measure (which itself generalizes the Plancherel measure) follows from considering gauge theories in four, five and six dimensions. These gauge theories arise from certain 5-brane configurations when the plane in which the brane lives is compactified to a cylinder and then to a torus. We also saw that $U(1)^N$ quiver gauge theories arising on a stack of $N$ M5-branes separated on a circle have a partition function which is exactly the partition function of periodic Schur process of period $N$. The correlation function of chiral operators in these gauge theories are also given by sum over young diagrams \cite{losev} and it would be interesting to relate these to the expectation value of random variables in the periodic Schur process. 

\section*{Acknowledgements}

A.I.  supported in part by the Higher Education Commission grant HEC-20-2518. AI gratefully acknowledges the support from Summer Simons workshop 2015 where part of this work was carried out.

\end{document}